\documentclass[preprint,authoryear]{elsarticle}

\usepackage[section] {placeins}
\usepackage{amsmath,amssymb,amstext,amscd,bm}
\journal{New Astronomy}
\def\astrobj#1{#1}

\usepackage{graphics}
\usepackage{graphicx}
\usepackage{epsfig}
\usepackage{amssymb}
\usepackage{amsthm}

\journal{New Astronomy}
\def\astrobj#1{#1}

\begin{document}
\begin{frontmatter}
\title{Analytical Studies of  \astrobj{NGC 2571}, \astrobj{NGC 6802}, \astrobj{Koposov~53} and \astrobj{Be 89}}
  \author[kayseri]{Ferhat F. \"{O}zeren}
      \author[kayseri]{\"{O}zg\"{u}n Arslan}
      \author[kayseri]{\.Ibrahim K\"{u}\c{c}\"{u}k\corref{cor1}}
      \ead{kucuk@erciyes.edu.tr}
      \author[kayseri]{\.Inci Akkaya Oralhan}
          \cortext[cor1]{Corresponding author}
\address[kayseri]{Erciyes University, Science Faculty, Astronomy and Space Science Department, 38039, Kayseri, Turkey}

\begin{abstract}
Astrophysical parameters (age, reddening, distance, radius, luminosity function, mass function, total mass, relaxation time and mass segregation) have been estimated for open clusters \astrobj{NGC 2571}, \astrobj{NGC 6802}, \astrobj{Koposov 53} and \astrobj{Be 89} by using the Two Micron All Sky Survey(2MASS) photometry. We analyse the color-magnitude diagrams and stellar radial density profiles. We have found that \astrobj{NGC 2571} is the youngest one having young main sequence stars while \astrobj{Be 89} is the oldest cluster. 
\end{abstract}

\end{frontmatter}

%\linenumbers

\section{Introduction}
Open Star Clusters (OCs) are very important tools in studying the formation and evolution of the Galactic disk. Determination of astrophysical and evolutionary parameters, such as distance, age, spatial density, core and limiting radii and relaxation time, are very important to construct and develop theoretical models on the Galactic structure. To understand how OCs evolve, it is important to take into account the external perturbation effects. In this paper, beside the astrophysical and the structural parameters we derived also the Mass Functions (MFs) for core, halo and overall regions of the four galactic open clusters, under consideration. The MF slope, $\chi$, relaxation time and evolutionary parameters are required in order to estimate the degree of mass segregation in different masses of cluster components \citep{bonatto2007}. In our analysis, we use 2MASS J and H bands for the spatial and photometric uniformities. The 2MASS Point Source Catalogue (PSC) is a uniform catalogue which allows to reach relatively faint magnitudes covering nearly all sky and defines a proper background situation for a cluster with large angular sizes \citep{bonatto2005}. The cluster parameters derived from 2MASS photometric data show some disagreements as compared with other studies on the same clusters. These disagreements arise since different analytical methods are used and the isochrone data sets differ when one takes into account the optical and near IR photometric systems. In order to determine more precise astrophysical and structural parameters  of the clusters, contamination effect of field stars might be reduced on their color-Magnitude Diagrams (CMDs) by selecting only stars whose magnitudes and colors are coherent with cluster. 

\section{The target open clusters}
The stellar radial density profiles of the clusters constructed with WEBDA\footnote{obswww.univie.ac.at/WEBDA-Mermilliod \& Paunzen \ (2003)} coordinates do not show definite peaks at central regions except for the radial density profiles (RDPs) of \astrobj{Koposov 53}. The number density, $r_{i}$, in the $i^{th}$ zone is calculated by using the 
\begin{figure}
\centering
\epsfig{file=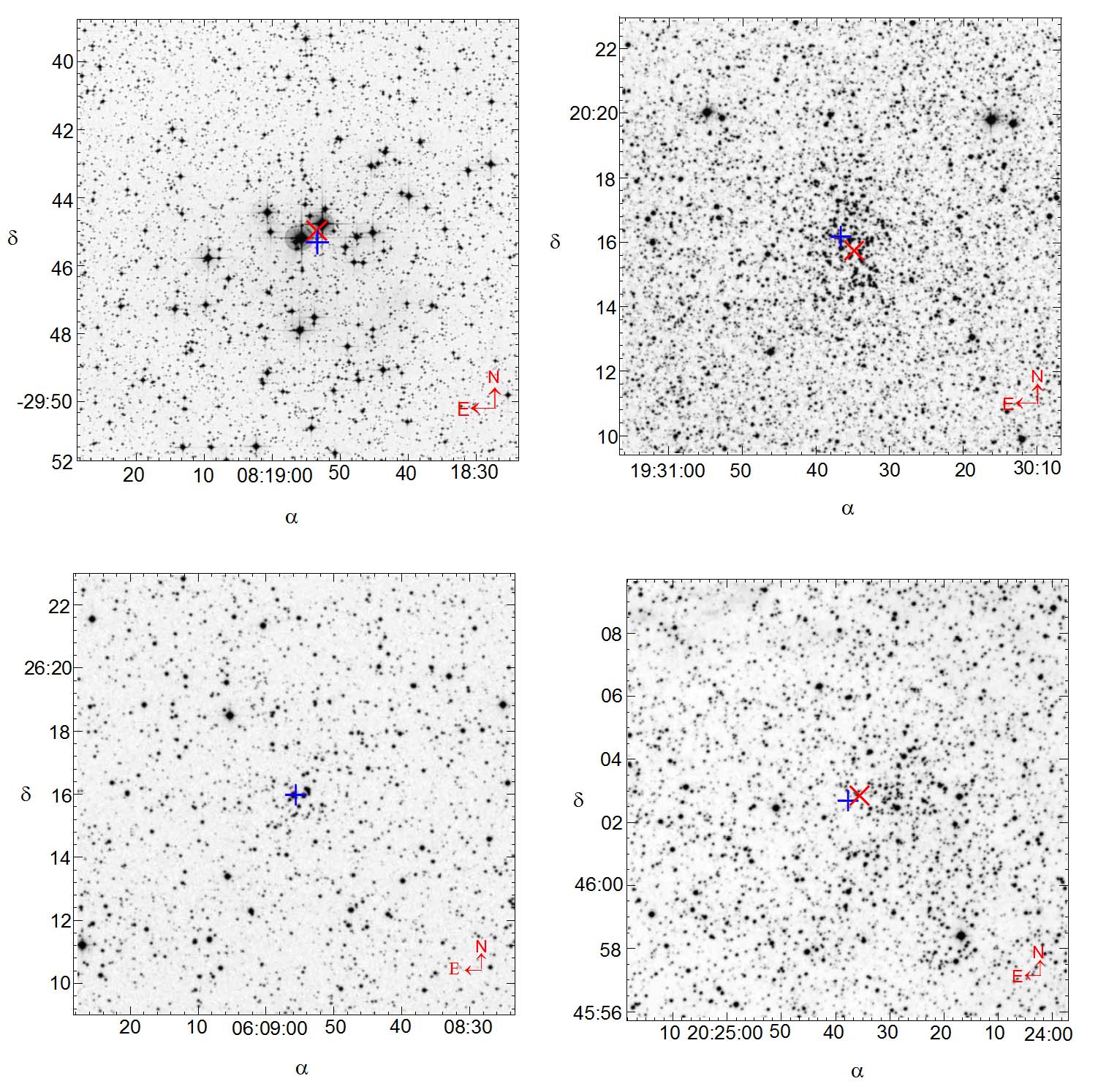, width=9cm, height=13.4cm}
\caption{Upper left panel: $10^{\prime} x 10^{\prime}$  DSS-I
 image of \astrobj{NGC 2571}. Upper right panel:  $10^{\prime} x 10^{\prime}$  DSS-I image of \astrobj{NGC 6802}. Bottom left panel: $10^{\prime} x 10^{\prime}$ DSS-I filled around \astrobj{Koposov 53}. Bottom right-hand panel: $10^{\prime} x 10^{\prime}$  DSS-I filled around \astrobj{Be 89}. The crosses show the new centers found in present study, and the pluses mark WEBDA central coordinates.}
        \label{figure1}
 \end{figure}
formula of $r_{i}$ =  $N_{i}$ /$A_{i}$. Where $N_{i}$ is the number of stars and $A_{i}$ is the area of the $i^{th}$ zone \citep{tadross2005a}. In this study the clusters under consideration are divided into central concentric circles with the adopted bin sizes ($0.8^{\prime}$, $0.5^{\prime}$, $0.5^{\prime}$ and $0.8^{\prime}$ for \astrobj{NGC 2571}, \astrobj{NGC 6802}, \astrobj{Koposov 53} and \astrobj{Be 89}, respectively). The purpose of this counting process is to determine the maximum central density of the clusters. The centers of the clusters are found by fitting Gaussian distribution function to the profiles of star counts. The maximum star densities of four clusters are displayed in Figure 2. The central coordinates of the clusters taken from WEBDA are summarized in Table 1 under Literature columns. The re-examined central coordinates and corresponding Galactic longitudes and latitudes are also given in Table 1. The DSS-I\footnote{Extracted from Leicester Database and Archive Servicehe (LEDAS) at \textit{http://www.ledas.uk/DSSimage/}}  images of \astrobj{NGC 2571}, \astrobj{NGC 6802}, \astrobj{Koposov 53} and \astrobj{Be 89} are displayed in Figure 1. 

\subsection{\astrobj{NGC 2571}}
The southern young open cluster \astrobj{NGC 2571} is a moderate young open cluster located in low absorption region associated with Vela Puppis \citep{giorgi2002} and projected in the 3rd Galactic quadrant. According to Trumpler and Ruprecht classification, it belongs to class I3p and II2p, respectively \citep{lindoff1968}. It appears to have not well defined evolutionary components such as main sequence and turn off due to field stars contamination and a small nuclear age (Figure 3). It has been studied previously by \citet{trumpler1930}, \citet{barkhatova1950}, \citet{lindoff1968}, \citet{claria1976}, and \citet{giorgi2002}. \citet{kilambi1978} and \citet{girardi2000} noticed the presence of a gap in the stellar distribution along the cluster$^\prime$s Main-Sequence (MS). The presence of this gap might mainly be explained by two different physical mechanisms: advanced convection processes in the stellar envelopes of intermediate age cluster members \citep{bohm1974} and the upper main sequence cluster stars resisted the gravitational contraction due to nuclear burning of $~^3He$ isotopes \citep{ulrich1971}. But this gap is not seen in our results because we didn't apply any decontamination procedure for the field stars in this cluster. Our distance modulus, reddening and age estimations might be considered to be almost independent of the gap, although decontamination of field star is important for understanding the real morphology of \astrobj{NGC 2571}$^\prime$s CMD. Different estimations taken into account for \astrobj{NGC 2571} are summarized as follows:

\begin{itemize}
\item[-]
 \citet{giorgi2002} suggested that roughly 30 \% of cluster members might be variable stars and the rests are candidates for metallic line stars, probably constituting a sequence of Am to Fm.
\item[-]
\citet{ahumada1995} claimed that the two brightest stars in the cluster region are blue stragglers and located far above the turn-off point if an age of 175 Myr is adopted.
\end{itemize}

\subsection{\astrobj{NGC 6802}}
The old open cluster \astrobj{NGC 6802} is located at the first Galactic quadrant where the cumulative effects of different reddenings and field stars contamination are more effective. CMD reveals fairly populous red giant clumps and well-defined main sequence 
\begin{figure}
\centering
\epsfig{file=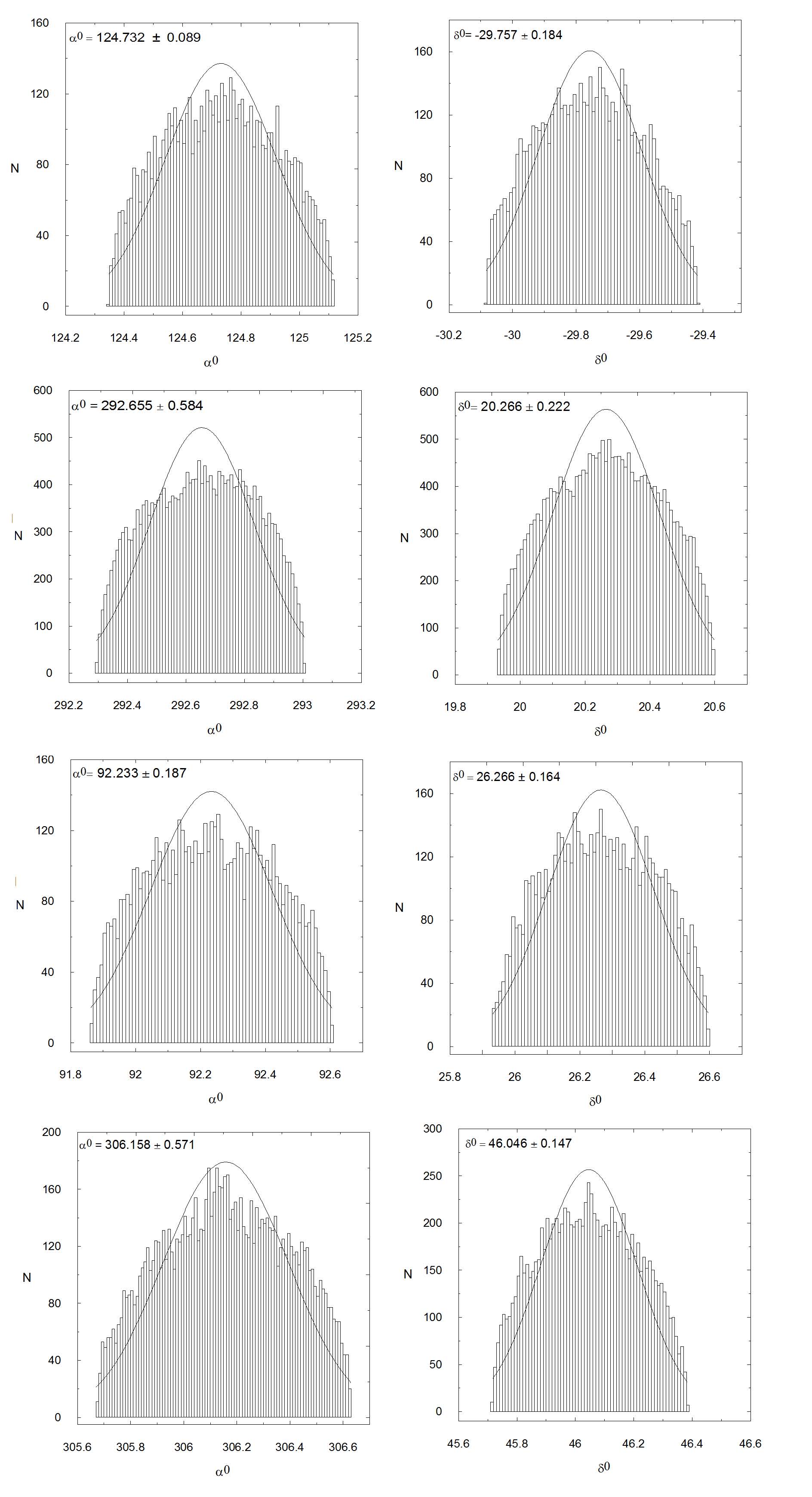, width=9cm}
\caption{Profiles of stellar counts across NGC 2571(first two panels), \astrobj{NGC 6802}(second two panels), \astrobj{Koposov 53}(third two panels) and \astrobj{Be 89}(last two panels) from top to bottom, respectively. The Gaussian fits have been applied. The center of symmetry about the peaks of $\alpha$ and $\delta$ is taken to be the position of the clusters$^\prime$ centers.}

        \label{figure2}
 \end{figure}
% Table 1
\begin{table*}
%\tiny
\centering
\caption{Sample coordinates.}
\renewcommand{\tabcolsep}{1mm}
\renewcommand{\arraystretch}{1.5}
\begin{tabular}{llllllllll}
\hline
&&\multicolumn{0}{c}{Literature}&&&&&\multicolumn{0}{c}{This paper}\\
\cline{2-5}
\cline{7-10}
\multicolumn{1}{c}{Cluster}&$\alpha(2000)$&$\delta(2000)$&$\ell$&b&&$\alpha(2000)$&$\delta(2000)$&$\ell$&b\\
&(h$\,  $m$\,  $s)&($^{\circ}$\, $^{\prime}$\, $^{\prime\prime}$)&($^{\circ}$)&($^{\circ}$)  &
&(h$\,  $m$\,  $s)&($^{\circ}$\, $^{\prime}$\, $^{\prime\prime}$)&($^{\circ}$)&($^{\circ}$)\\
\astrobj{NGC$\,$2571} &08 18 56&$-$29 45 00&249.106&3.532& &08 18 56&$-$29 45 33&249.111&3.529\\
\astrobj{NGC$\,$6802} &19 30 35&+20 15 42 &55.326&0.918 & &19 30 37 &+20 15 56&55.333&0.912\\
\astrobj{Koposov$\,$53} &06 08 56&+26 15 55&184.901&3.131& &06 08 56&+26 15 55&184.901&3.131\\
\astrobj{Be$\,$89} &20 24 36 &+46 03 00&83.169&4.822& &20 24 38&+46 02 45&83. 159&4.815\\
\hline
\end{tabular}
\end{table*}
\begin{figure}
\centering
\epsfig{file=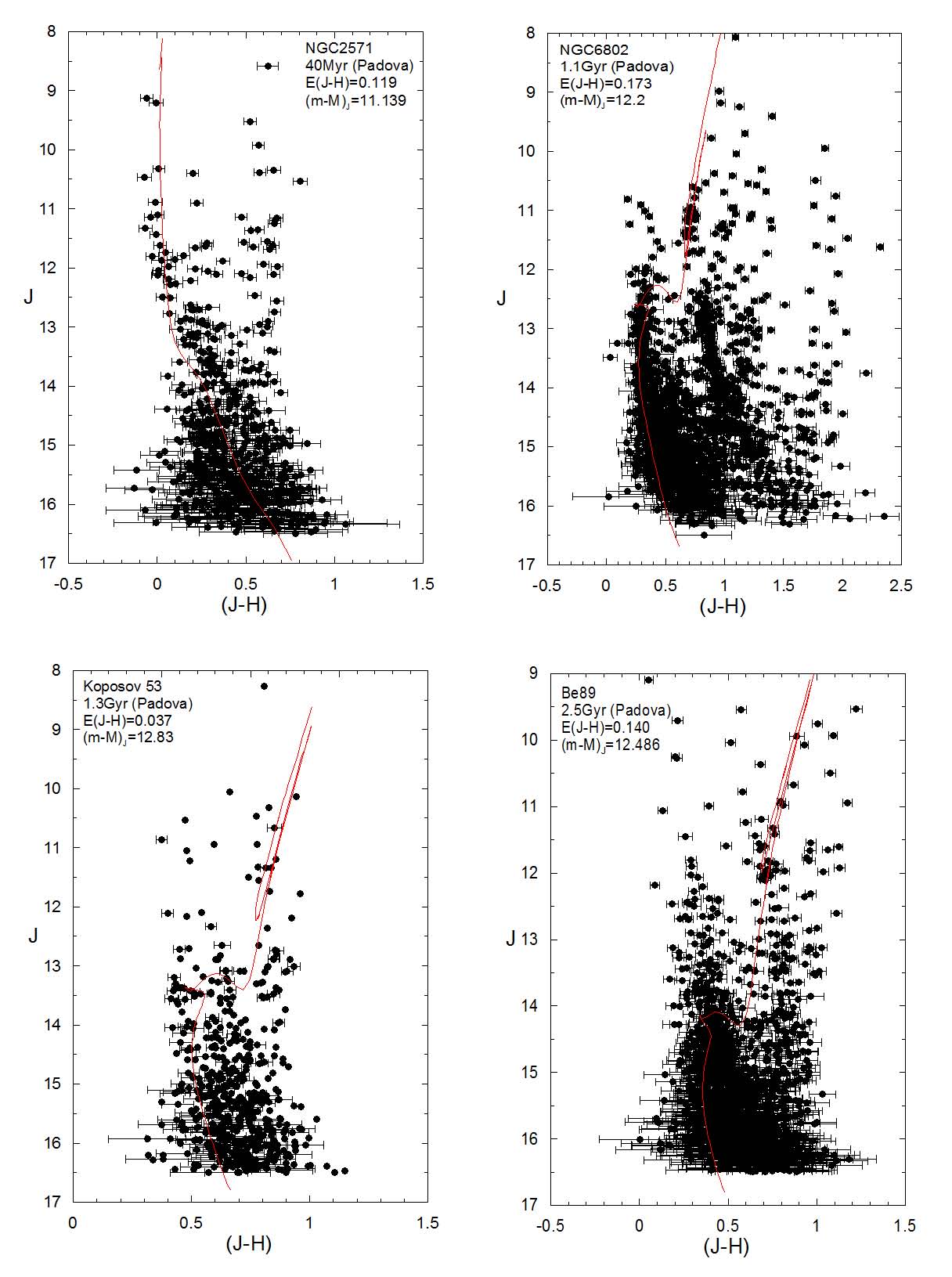, width=13cm}
\caption{Observed $J x (J-H)$ CMDs extracted from the field region of $20^{\prime}$ for clusters under consideration. The solid lines represent the fitted Padova solar isochrones. The CMDs have been constrained using limiting radius values which are $5^{\prime}.7$, $5^{\prime}$, $6^{\prime}$ and $6^{\prime}.5$, for  \astrobj{NGC 2571}, \astrobj{NGC 6802}, \astrobj{Koposov 53}  and \astrobj{Be89}, respectively}
    \label{figure3}
 \end{figure}
(the upper right in Figure 3). Because of its large distance from the Sun, low galactic latitude and the embedded spatial distribution, it is difficult to determine astrophysical parameters more accurately. The low Galactic location may also lead to obstruct understanding of the realistic mass distribution of the cluster. Different estimations taken into account for \astrobj{NGC 6802} are as follows:
\begin{itemize}
\item[-]
 \citet{dutra2000} compared reddening estimations of \astrobj{NGC 6802}. A far - infrared reddening value of $E(B-V)_{FIR}$ derived from DIRBE/IRAS 100 $\mu$m dust emission map of \citet{schlegel1998} is used. They suggested a far-infrared reddening of $E(B-V)_{FIR} = 4.3$, which is quite different from stellar ones due to the variance of dust distribution along line of sight of the cluster.
\end{itemize}

\subsection{\astrobj{Koposov 53}}
\astrobj{Koposov 53}, is projected against the Galactic anti-center and distinctly visible from the surrounding region \citep{koposov2008}. The distribution of field stars might be accepted as near homogeneous.  The cluster is located at $3^{th}$ quadrant where the photometric data are affected by differential reddening from the interstellar dust. \astrobj{Koposov 53} has a moderately distinguishable extended main sequence, and most likely, a few red giant stars. These are not highly separated from the field region (Figure 3). The cluster main sequence might be partly explained by the combined effect of unsolved binary systems and contamination of field stars whose color-magnitude properties are similar to cluster members. 

\subsection{\astrobj{Be 89}}
The northern open cluster \astrobj{Be 89} is located at a high latitude in the first Galactic quadrant. The Galactic location is consistent with its old age, which is very important for understanding the history of Galactic disc. \astrobj{Be 89}, belonging to III1p subclass of Trumpler, is not particulary well studied object due to large differential reddening. Both of strong reddening and field stars contamination cause separation in CMD, especially in the main sequence region.  The internal and external dynamical effects as well as the photometric ones mentioned before, lead to an important decrease of spatial discrimination of \astrobj{Be 89}. The prominent characteristic feature of CMD of \astrobj{Be 89} is the presence of a broaden main sequence. This main sequence is heavily contaminated by field stars and have a well defined giant branch as compared with other samples taken into account in this work (Figure 3). 

\section{The 2MASS photometry}
Depending on the Near-IR JH${K_{s}}$ 2MASS data, deep stellar analyses of four open clusters have been presented. 2MASS is uniformly scanning the entire sky in three near IR bands J (1.65 $\mu$m), H (1.65 $\mu$m) and $K_{s}$  (2.17 $\mu$m). On the 2MASS scale, the completeness limits are found to be 16.5, 15.8 and 15.2 mag for J, H and $K_{s}$ bands, respectively \citep{tadross2005a}. The photometric data of 2MASS not only allow the construction of relatively well defined CM diagrams of the clusters, but also permit a more reliable determination of astrophysical parameters. Additionally, the relatively low interstellar extinction ratios in near infrared wavelengths provide us an opportunity of comprehensive research to investigate spiral arm structures where most of open clusters are intensively situated. In this paper, we used extraction areas having radius of $20^{\prime}$ which are larger than estimated limiting radius of the clusters. Because of the weak contrast between the cluster and background field density, some inaccurate statistical results may be produced beyond the real limit of cluster borders \citep{tadross2005b}. As photometric quality constraints, we restricted the extraction range between 8$<$J$<$16 in the J magnitude. For reddening transformations we used the relation $A_{J}/{A_{V}}=0.276$, $A_{H}/{A_{V}}=0.176$, $A_{K_{s}}/{A_{V}}=0.118$, $A_{J}=2.76\times{E(J-H)}$, and $E(J-H)=0.33\times{E(B-V)}$ \citep{dutra2002}, assuming a constant total to selective absorption ratio $R_{V}=3.1$.

\section{Results of astrophysical parameters}
Depending on the photometric and statistical qualities of the 2MASS photometry, the astrophysical parameters of clusters, such as age, reddening, distance modulus, can be determined by fitting the isochrones to the cluster CMDs. To do this, we applied several fittings on the $J-(J-H)$ CMDs of the clusters by using \citep{girardi2000} Padova isochrones on the solar metallicity. It is worth mentioning that the assumptions of solar metallicity are quite adequate for young and intermediate-age open clusters which are projected close to against the Galactic disc, as in our samples. 
We preferred fitting the isochrones to the redder envelopes of points matching the stellar components of the four clusters, because of the probable main sequence color extension effect. We note that, there are some disagreements in the astrophysical parameters, especially for reddening and age, with previous works. These disagreements come from the locations of the our samples along the Galactic plane where the field stars contamination may be important and the characteristic features of the different photometric systems are used for analyses. In most cases, it is difficult to obtain accurate determination of the astrophysical parameters due to the weak contrast between clusters and field stars. 

\textbf{The reddening results:} Reddening determination is one of the major steps in the cluster compilation. The $E(J-H)$, color excess parameters derived from the isochrone fits are converted to $E(B-V)$ values. Our $E(B-V)$ reddening estimations are 0.36 for \astrobj{NGC 2571}, 0.52 for \astrobj{NGC 6802}, 0.11 for  \astrobj{Koposov} 53 and 0.42 for \astrobj{Be 89}. The relatively lower reddening value $E(B-V)$=0.11 of \astrobj{Koposov 53} can be explained by its Galactic location where the differential reddening is expected to decrease for the Galactic anti-center. Thus, near-infrared surveys are very useful for the investigation of the clusters like \astrobj{Koposov 53}. It is relatively less affected by high reddening from Galactic plane. We compared the reddening values of the four clusters with the dust maps of \citet{schlegel1998}(hereafter SFD). These are based on the COBE/DIRBE and IRAS/ISSA maps, and take into account the dust absorption all way to infinity \citet{gunes2012}. Using the method of \citet{bonafacio2000}, and our distance values we determined the corrected SFD reddening estimations for our clusters. The reddening values are $E(B - V)_{A}$ = 0.18, 0.72, 0.28 and 0.61 for  \astrobj{NGC 2571}, \astrobj{NGC 6802}, \astrobj{Koposov 53}  and \astrobj{Be 89}, respectively. The reddening estimations obtained from 2MASS are close to the SFD values, except for \astrobj{NGC 2571}. This estimation is higher than $E(B - V)_{A}$ value of SFD, and probably because of the young age of the cluster. The differences between our reddening estimations and SFD results are quite acceptable for these clusters. Because, the SFD reddening results with relatively low spatial resolution are obtained along the line of sight.

\textbf{The distance modulus:} We found absolute distance modulus $(V-M_{v})_{o}$ as 1.69, 2.16, 3.50 and 3.14 for \astrobj{NGC 2571}, \astrobj{NGC 6802}, \astrobj{Koposov 53} and \astrobj{Be 89}, respectively. Under the assumption of R$_\odot$ = 7.2 $\pm$ 0.03 kpc of \citet{bica2006} which is based on updated parameters of globular clusters, the estimated distances from the Galactic center, $R_{GC}$, are found to be 7.98, 6.25, 10.7 and 7.52 kpcs for \astrobj{NGC 2571}, \astrobj{NGC 6802}, \astrobj{Koposov 53} and \astrobj{Be 89}, respectively. \astrobj{NGC 6802} is inside the Solar Circle, whereas \astrobj{NGC 2571}, \astrobj{Koposov 53} and \astrobj{Be 89} are outside. The estimations of distance modulus and distances (heliocentric and Galactocentric), which are also given in columns 3, 4 and 5 of Table 2, are in agreement with previous studies.

\textbf{The age results:} Cluster ages are derived by means of solar-metallicity Padova isochrones \citet{girardi2000}. As seen on their CMDs the density of cluster members are mostly comparable or even less than field stars$^\prime$ population. Because of strong field stars contamination, it is not possible to completely separate all field stars from cluster members (Figure 3). Our age estimations, which are also given in column 6 of Table 2 are 40 Myr, 1.1 Gyr, 1.3 Gyr and 2.5 Gyr for \astrobj{NGC 2571}, \astrobj{NGC 6802}, \astrobj{Koposov 53}, and \astrobj{Be 89}, respectively. It is worth mentioning that there are some differences between our age values and the previous estimations. This situation might be mainly explained by the theoretical differences in adopted isochrones.
% Table 2
\begin{table*}
\centering
%\tiny
\caption{Comparison of the fundamental astrophysical parameters of \astrobj{NGC 2571}, \astrobj{NGC 6802}, \astrobj{Koposov 53}, \astrobj{Be 89} with those given in literature. Columns.~2$-$7 represent the reddening, true distance modulus, heliocentric distance, Galactocentric distance, age and size, respectively.}
\renewcommand{\tabcolsep}{2mm}
\renewcommand{\arraystretch}{1.5}
\begin{tabular}{ccccccccc}
\hline
Cluster&$E(B-V)$& (V-M$_{v})_{0}$& d(kpc) & R$_{GC}$ & A(Gyr) &Size&Photometry&Reference \\
\hline
\multicolumn{ 1}{c}{\astrobj{NGC 2571}}&0.36&11.14&1.69&7.98&0.004&5.7$^{\prime}$&2MASS&This paper \\
\multicolumn{ 1}{c}{}&0.32&-&-&-&-&12$^{\prime}$&UBV&[1]\\
\multicolumn{ 1}{c}{}&0.52&11.67&2&-&0.027&-&UBV&[2] \\
\multicolumn{ 1}{c}{}&0.10&10.41&1.21&-&0.175&-&UBV&[3] \\
\multicolumn{ 1}{c}{}&0.10&10.69&1.38&-&0.005&-&UBVI&[4] \\
\hline
\multicolumn{ 1}{c}{\astrobj{NGC 6802}}&0.52&11.67&2.16&6.25&1.1&5$^{\prime}$&2MASS&This paper\\
\multicolumn{ 1}{c}{}&0.94&10.78&1.43&-&1&-&BVRI&[5] \\
\multicolumn{ 1}{c}{}&0.89&11.64&2.13&-&0.5&-&$\Delta{a}$&[6] \\
\multicolumn{ 1}{c}{}&0.84&11.25&1.77&7.14&0.96&-&BVRI&[7] \\
\hline
\multicolumn{ 1}{c}{\astrobj{Koposov 53}}&0.11&12.72&3.50&10.70&1.3&$6^{\prime}$&2MASS&This paper\\ 
\multicolumn{ 1}{c}{}&0.34&12.52&3.2&-&0.32&$3^{\prime}$&2MASS&[8] \\
\multicolumn{ 1}{c}{}&0.41&13.62&5.3&-&0.089&$4.2^{\prime}$&UBVI&[9] \\
\hline
\multicolumn{ 1}{c}{\astrobj{Be 89}}&0.42&12.48&3.14&7.52&2.5&6.5$^{\prime}$&2MASS&This paper\\ 
\multicolumn{ 1}{c}{}&-&12.38&3&-&1&-&2MASS&[10] \\
\multicolumn{ 1}{c}{}&1.05&11.54&2.04&-&1-1.12&-&BVI&[11] \\
\multicolumn{ 1}{c}{}&0.6&11.90&2.40&8.55&0.38&-&UBVRI&[12]\\
\hline
\end{tabular}
\\
%{\tiny
{$[1]$: Lindoff et al. (1968), $[2]$: Claria et al. (1976), $[3]$: Kilambi et al. (1978),$[4]$: Giorgi et al. (2002), $[5]$: Sirbaugh et al. (1995),$[6]$: Netopil et al. (2007),$[7]$: Janes et al. (2007),
%\citep{janes2007},
$[8]$: Koposov et al. (2008), $[9]$: Yadav et al. (2011), $[10]$: Tadross (2008), $[11]$: Subramaniam et al. (2010), $[12]$: Akkaya et al. (2010)  \\
}
\end{table*} 
% Table 3
\begin{table*}
\centering
{\scriptsize
\caption{Structural parameters of the four clusters.}
\begin{tabular}{lllllll}
\hline
Cluster&1$^{\prime}$(pc)&$\sigma_{bg}(star$\,$ pc^{2}$)&$\sigma_{ok}(star$\,$ pc^{2}$)&R$_{lim}(pc)$&R$_{core}(pc)$&$R_{t}(pc)$\\
\hline
\astrobj{NGC$\,$2571}   &0.49&23.41&20.91&2.80&0.84&10.71\\
\astrobj{NGC$\,$6802}   &0.63&73.03&46.44&3.15&0.54&12.77 \\
\astrobj{Koposov$\,$53} &1.02&14.02&4.16&6.10&0.73&12.55 \\
\astrobj{Be$\,$89}      &0.91&7.77&9.11&5.94&2.56&13.06\\
\hline
\end{tabular}
}
\begin{list}{ Notes.}
\footnotesize
\item Col. 2: arcmin to parsec scale. 
\end{list}
\end{table*}
\subsection{Limiting radius}
The limiting radius of a cluster can be described with an observational border which depends on the spatial distribution of stars in cluster and number density of cluster members and degree of field-star contamination. We determined RDPs of the clusters by examining the stellar spatial distributions. For this purpose, the projected stellar density of the objects were determined by counting stars within concentric circles (zones) around the  determined cluster centers. To avoid over-sampling near the center and under-sampling for large radii, the RDPs were built by counting stars in concentric rings of increasing width with distance to the center. The number and width of the rings are adjusted to background and/or foreground stellar density to present adequate spatial resolution. It might be claimed that most of the stars in the inner concentric rings are quite likely members, whereas the external rings are more intensely contaminated by field stars. Consequently, we estimated the limiting radii with their uncertainties by considering the fluctuations of the RDPs with respect to stellar background, and the cluster radii correspond to distances from the centers where RDPs show distinctive decrease. Since the accurate estimation of limiting radius depends directly on the statistical fluctuation degrees of RDP, which is driven by background level, it is expected that measurements of poor open cluster populations, particularly those affected by strong field-star contamination, will be included large uncertainties. Our limiting radius estimations, $R_{lim}$, are $5^{\prime}.7$, $5^{\prime}$, $6^{\prime}$ and $6^{\prime}.5$ arcmin for  \astrobj{NGC 2571}, \astrobj{NGC 6802}, \astrobj{Koposov 53}  and \astrobj{Be 89}, respectively, as seen in Table 3. Additionally, the limiting radius values have been converted to parsec unit in order to compare with tidal radii of the clusters.

\subsection{Structural parameters}
Structural parameters were derived by fitting the observed RDPs with two parameter King surface density profile, which describes the intermediate and central regions of normal globular clusters \citep{king1966}. Using the King profile fitting we derived structural parameters of the clusters such as central density of stars $(\sigma_{ok})$ , stellar background level $(\sigma_{bg})$, and core radius $R_{core}$. Core and radii relationship is intimately related to the cluster survival/dissociation rates. Both kinds of radii present a similar dependence on age, where part of clusters expand with time, while some seem to shrink. To determine the structural parameters of the clusters which are listed in Table 3, we adopt two parameters distribution function $\sigma(R) = \sigma_{bg} + \sigma_0/(1+(R/R_{core})^2)$.
The RDPs of stellar components of our samples such as main sequence and giant branch should highly obey to King Profile Law, which describes the structural features of central and external regions of a star cluster. Because these objects are not dense and gravitationally bounded systems, like globular clusters, some difficulties in the modelling arise. Nevertheless, the King$^\prime$s profile analytically represent an adequate stellar RDPs of the four clusters from the external parts (the halo) to the center (the core) (Figure 4).

\subsection{Tidal radius}
Determination of the tidal radius of a cluster might be properly done by the spatial coverage and uniformity of 2MASS photometry, which  allows one to obtain reliable data relating to spatial distribution of cluster stars. The tidal radii of open clusters depend on both combined effects of galactic tidal fields and subsequent internal relaxation - dynamical evolution of clusters \citep{allen1988}. 
\begin{figure}
\centering
\epsfig{file=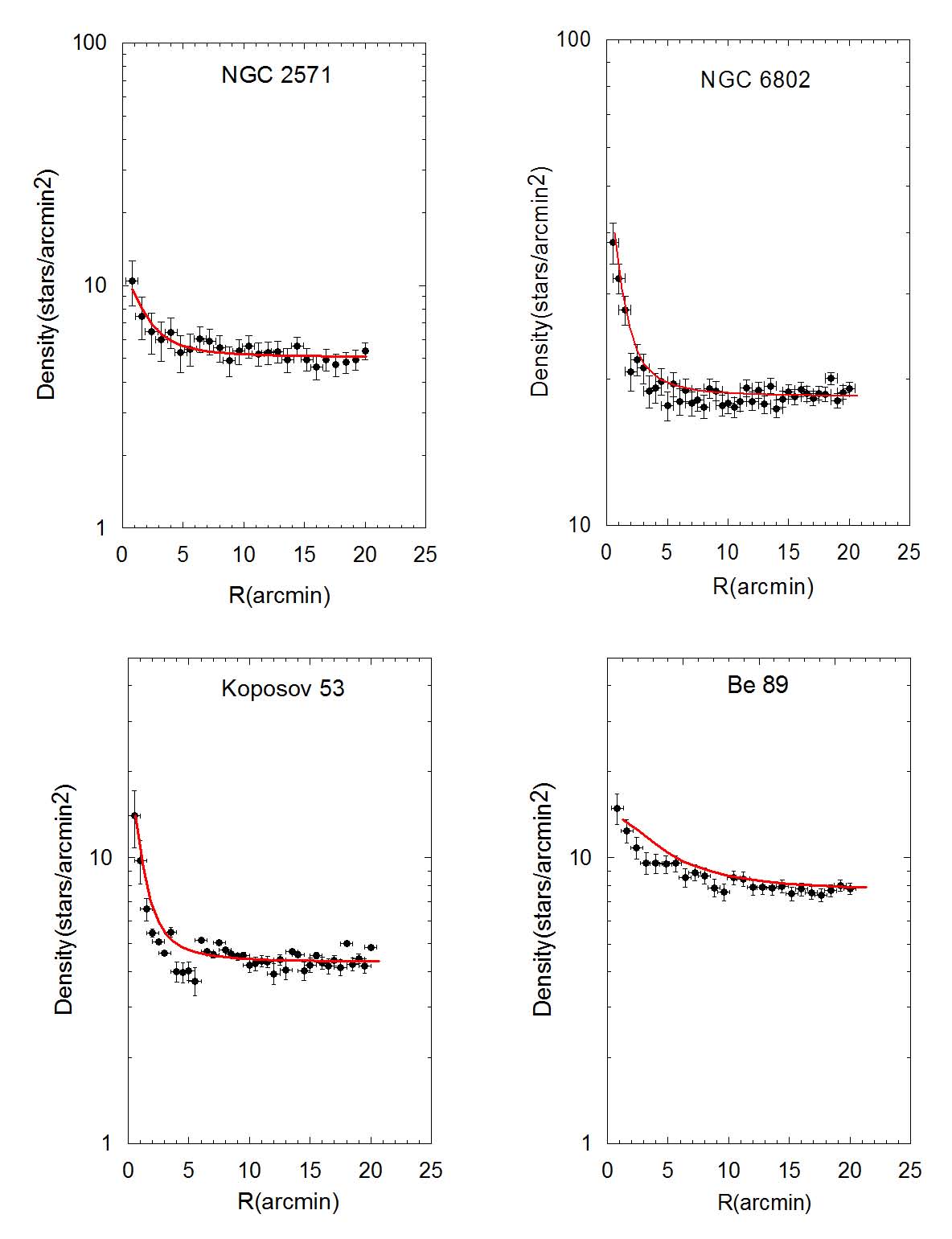, width=13cm}
\caption{Stellar RDPs (filled circles) of the clusters which have been built from 2MASS photometry. Solid lines show the best-fit King profiles. The lengths of the vertical and horizontal error-bars consist with Poisson distribution.}
    \label{figure4}
 \end{figure}
The tidal Radius $R_{t}$ of a cluster can be estimated by using the formula of \citep{kim2000}:                 	                                                                                          
\begin{equation}
R_{t}=(\frac{m}{2M_{g}})^{\frac{1}{3}}R_{g}
\end{equation}
Where $m$ is the mass of the cluster (our total mass estimations are given in Section 6), and $M_{g}$ is the Galactic mass inside the Galactocentric radius of cluster $R_{g}$ \citep{tadross2005c}. From \citep{genzel1987}, $M_{g}$ can be calculated as:                                                                
\begin{equation}
M_{g}=2 \times 10^{8} \times (\frac{R_{t}}{30 pc})^{1.2}
\end{equation}	
According Equation (1), our tidal radius, $R_{t}$, estimations are 10.71, 12.77, 12.55 and 13.06 pcs for \astrobj{NGC 2571}, 
\astrobj{NGC 6802}, \astrobj{Koposov 53} and \astrobj{Be 89}, respectively.

\section{Luminosity functions}
The measurements of the number of stars in a cluster with a given color and magnitude ranges are very important to understand the characteristic properties of the evolutionary stages of these objects. For this purpose, we obtained the Luminosity Functions (LFs) of the four clusters by summing up the J band luminosities of all stars within the determined limiting radii. Before building
\begin{figure}
\centering
\epsfig{file=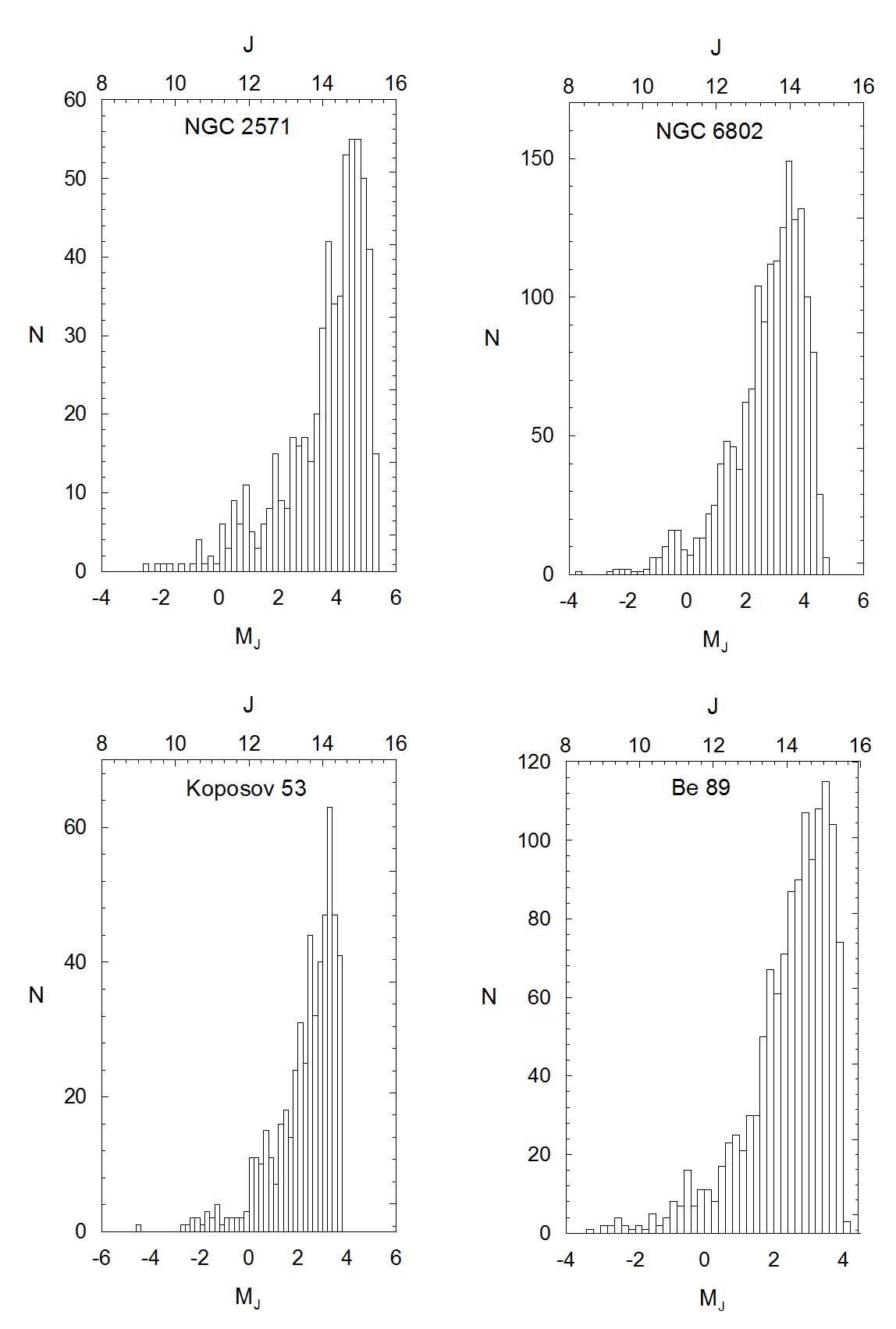, width=12cm}
\caption{Spatial distribution of LFs for the clusters in terms of the absolute magnitude $M_{J}$. The apparent J magnitude scales have been placed along the upper axes of the panels. The absolute peaks lie at 4.50, 3.47, 3.30 and 3.50 mag for J band, respectively.}
    \label{figure5}
 \end{figure}
the LFs, we converted the apparent J band magnitudes of possible member stars into the absolute magnitude values using the distance moduli of the samples. We constructed the histogram sizes of LFs to include a reasonable number of stars in each absolute J magnitude bins for the best counting statistics. The total LFs of the cluster are found to be -4.91, -6.45, -6.03 and -6.58 mag for \astrobj{NGC 2571}, \astrobj{NGC 6802}, \astrobj{Koposov 53} and \astrobj{Be 89}, respectively (Figure 5).

\section{Total masses, mass functions and dynamical states}
In this work, we estimated MFs of the four clusters using the theoretical evolutionary tracks and their isochrones with different ages. The masses of possible cluster members were derived from the polynomial expression developed by \citep{girardi2000} with  solar metallicity. To obtain mass functions we follow the same process we have used previously for RDPs. We also estimated
\begin{figure}
\centering
\epsfig{file=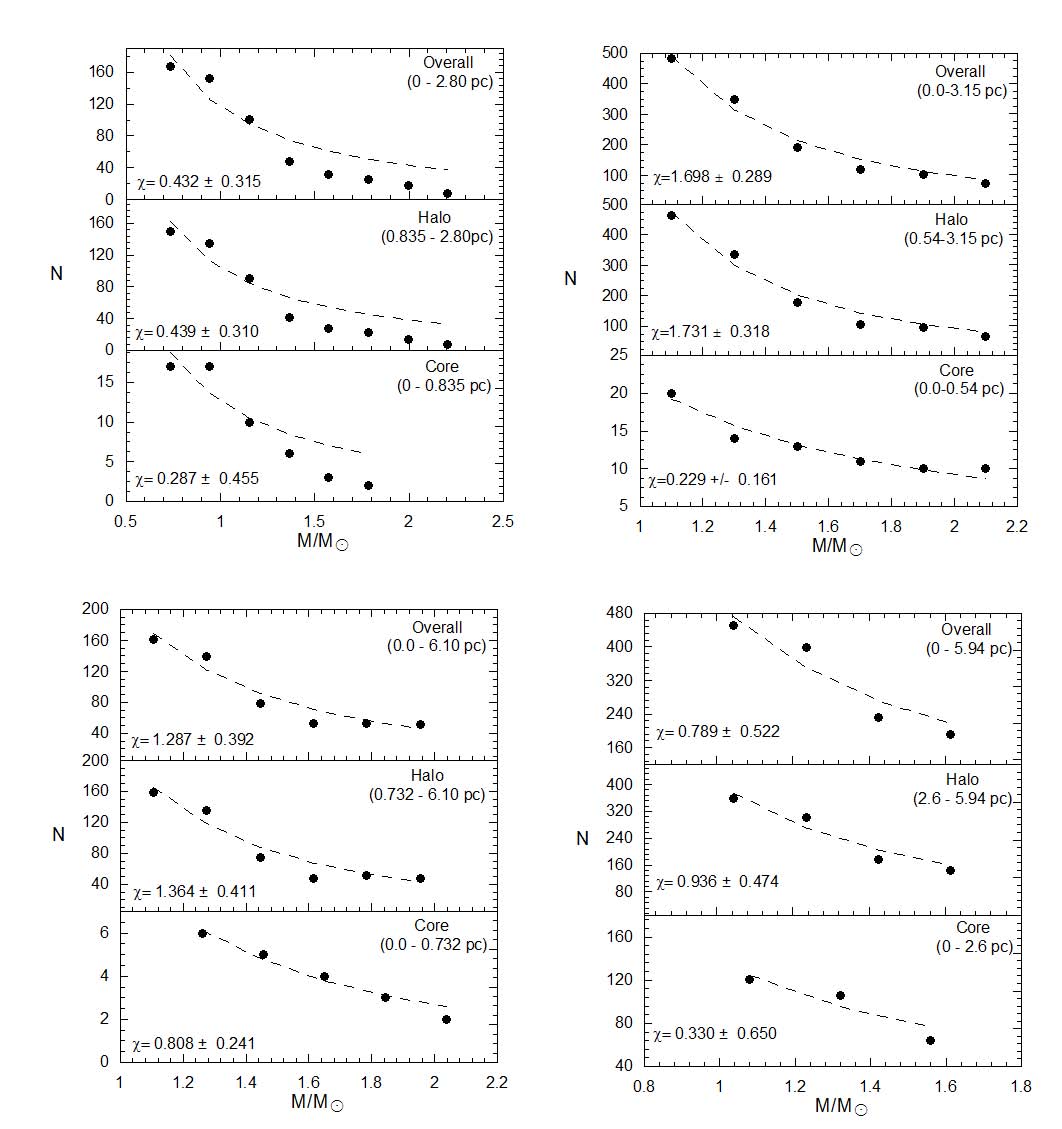, width=15cm}
\caption{Overall (top panels), halo (middle) and core (bottom) MFs of \astrobj{NGC 2571}( upper left panel), \astrobj{NGC 6802}(upper right panel), \astrobj{Koposov 53} (bottom left panel) and \astrobj{Be 89} (bottom right panel). Dashed curves present best fits for the function $\phi(m)\propto{m}^{-(1+\chi)}$.}
    \label{figure6}
 \end{figure}
total masses of the clusters in solar unit by multiplying the number of stars in each mass bin. Our total mass estimations are 785, 2067, 746 and 1585 M$_\odot$ for \astrobj{NGC 2571}, \astrobj{NGC 6802}, \astrobj{Koposov 53} and \astrobj{Be 89}, respectively. We emphasize that the total mass estimation should be taken as upper limit. Because, the field-star contamination effects are more efficient in crowded regions. 
% Table 4
\begin{table*}
\centering
{\scriptsize
\caption{Parameters derived for \astrobj{NGC 2571}, \astrobj{NGC 6802}, \astrobj{Koposov 53}  and \astrobj{Be 89}. 
}
\begin{tabular}{ccrcc}
\hline
Cluster                         &Region(pc)     &N$^{*}$(Stars)&$\chi$ \\ \hline
\multicolumn{ 1}{l}{}   &0 - 2.80       &598           &0.432$\pm$0.315\\
\multicolumn{ 1}{l}{\astrobj{NGC$\,$ 2571}}           &0.835 - 2.80   &529           &0.439$\pm$0.310\\
\multicolumn{ 1}{l}{}           &0 - 0.835      &69            &0.287$\pm$0.435\\ \hline
\multicolumn{ 1}{l}{}  &0 - 3.15       &1625          &1.698$\pm$0.289\\
\multicolumn{ 1}{l}{\astrobj{NGC$\,$ 6802}}           &0.54 - 3.15    &1544          &1.731$\pm$0.318\\
\multicolumn{ 1}{l}{}           &0 - 0.54       &81            &0.229$\pm$0.181\\ \hline
\multicolumn{ 1}{l}{}   &0 - 6.10    &536           &1.287$\pm$0.342\\
\multicolumn{ 1}{l}{\astrobj{Koposov$\,$53}}              &0.732 - 6.10&515           &1.364$\pm$0.411\\
\multicolumn{ 1}{l}{}              &0 - 0.732   &21            &0.808$\pm$0.241\\ \hline
\multicolumn{ 1}{l}{}        &0 - 5.94    &1272          &0.789$\pm$0.522\\
\multicolumn{ 1}{l}{\astrobj{Be$\,$89}}              &2.6 - 5.94  &981           &0.936$\pm$0.474\\
\multicolumn{ 1}{l}{}              &0 - 2.6     &291           &0.303$\pm$0.650\\ \hline
\end{tabular}
}
\begin{list}{ Notes.}
\tiny
\item Col.1: the distance from the core. Col 2: possible cluster stars$^\prime$ number for the regions in Col 1. Col 3: present MF slopes $(\chi)$$\,$$\,$for different clusters regions.	 
\end{list}
\end{table*}
It is also worth mentioning that binary stars, in our samples, effected the color-magnitude distribution. Thus, the degree of accuracy of MFs have been constrained by lower masses secondary stars in binary systems with unequal mass components.

\subsection{Mass functions}
Figure 6 shows the core, halo and overall MFs of the four clusters together with the fits, which has been calculated by using the function $\phi(m)\propto{m}^{-(1+\chi)}$. The MFs slopes are given in Table 4. The slopes are evidently flat in cores and quite steep in outer regions. 
% Table 5
\begin{table*}
\centering
{\scriptsize
\caption{Number and total mass estimations of main sequence and evolved star. 
}
\begin{tabular}{ccrr}
\hline
Cluster&Star$\,$type&N$^{*}$(Stars)&M$_{\odot}$\\
\hline
&&& \\
\multicolumn{ 1}{l}{\astrobj{NGC 2571}}    &MS&549&571.54\\
\multicolumn{ 1}{l}{}             &TO$\,$+$\,$GB&49&213.47\\
\hline
&&& \\
\multicolumn{ 1}{l}{\astrobj{NGC$\,$6802}}     &MS&1479&1780.38\\
\multicolumn{ 1}{l}{}              &TO$\,$+$\,$GB&140&280.85\\
\hline
&&& \\
\multicolumn{ 1}{l}{\astrobj{Koposov$\,$53}}    &MS&449&579.70\\
\multicolumn{ 1}{l}{}               &TO$\,$+$\,$GB&87&166.01\\
\hline
&&& \\
\multicolumn{ 1}{l}{\astrobj{Be$\,$89}}          &MS&1113&1340.71\\
\multicolumn{ 1}{l}{}                 &TO$\,$+$\,$GB&159&244.20\\
\hline
\end{tabular}
}
\begin{list}{ Notes.}
\tiny
\item Col. 1: Main-Sequence$\,$(MS), Turn-Off$\,$(TO) and Giant Branch$\,$(GB).
\end{list}
\end{table*}
% Table 6
\begin{table*}
\centering
{\scriptsize
\caption{Relaxation time and evolutionary parameters of the clusters, derived from Binney \& Tremanine (1987) method for core and overall regions.}
\begin{tabular}{crrcr}
\hline
Cluster&$T_{R}$\,$(Myr)$&$\tau$&$T_{R}$\,$(Myr)$&$\tau$\\
\hline
&&&& \\
\astrobj{NGC$\,$2571}   &0.127&31.44&24.5&1.63\\
\astrobj{NGC$\,$6802}   &0.093&1179.79&64.7&17.01\\
\astrobj{Koposov$\,$53} &0.047&2757.50&48.6&26.72\\
\astrobj{Be$\,$89}      &12.3&203.60&98.8&25.30\\
\hline
\end{tabular}
}
\end{table*}
% Table 7
\begin{table*} 
\centering
{\scriptsize
\caption{The comparison of different overall T$_{R}$ results, derived from Binney \& Tremanine (1987) and Spitzer \& Hart (1971).}
\begin{tabular}{ccc}
\hline
             &Binney$\,$(1987)&Spitzer$\,$(1971)\\
Cluster      &T$_{R} \,$(Myr)          &T$_{R} \,$(Myr)\\ \hline
\astrobj{NGC$\,$ 2571}   &24.5                  &22.0\\
\astrobj{NGC$\,$ 6802}   &64.7                  &33.5\\
\astrobj{Koposov$\,$ 53} &48.6                  &44.3\\
\astrobj{Be$\,$ 89}      &98.8                  &84.1\\ \hline
\end{tabular}
}
\end{table*}
These variations of MFs slopes from the internal parts to the halo of the clusters might be interpreted as follows: The core low mass member stars are forced to move towards external parts while massive stars accumulate in the core, because of mass segregation which has been driven by dynamical evolution process. In other words, relatively young massive cluster member stars have gradually condensed into central regions with raising negative binding energy, whereas old member stars have tended to populate towards cluster halo, decreasing their binding energy. The explicit spatial variance of MF slopes in our samples may be due to a large amount of massive member stars initially formed close to central region and stayed there. Or, it may be the result of dynamical processes related to mass segregation which has led to significant deviation of mass distribution from spatial homogeneity. The results are displayed in Table 4, 5, 6 $\&$ 7 and discussed below. 

\textbf{\astrobj{NGC 2571}:}  There is an evident variation of MF slopes. In the core we found $\chi=0.287 \pm 0455$, in the halo it becomes $\chi=0.439 \pm 0.310$. Overall slope is found to be $\chi=0.432 \pm 0.315$. The variation is quite large from the core to the halo of the cluster \astrobj{NGC 2571}, due to the large scale mass segregation. One can say that \astrobj{NGC 2571} is relatively young open cluster having less efficient nuclear and dynamical evolution processes.

\textbf{\astrobj{NGC 6802} :} The MF slopes $\chi$ of \astrobj{NGC 6802} are $0.229 \pm 0111$, $1.731 \pm 00318$ and $1.698 \pm 00289$ for the core, the halo and the overall regions, respectively. The overall MF slope of this cluster is different from the standard initial mass function (IMF) value of $\chi$=1.35 \citep{salpeter1955}. \astrobj{NGC 6802} has quite large overall MF slope compared with its core value probably because of combined effects of mass segregation and gradual core collapse phase.

\textbf{\astrobj{Koposov 53} :} The MF slope is flat ($\chi=0.880 \pm 0.241$) in the core, while more steep in the halo ($\chi=1.364 \pm  0.411$). Overall value is $\chi=1.287 \pm  0.392$ which is different from Salpeter$^\prime$s value. But, the MF slope of the halo is quite close to standard value. Additionally, the relatively steady variation of MF slopes from the core to outer regions implies that a mild mass segregation and low mass stars evaporation processes have occurred in \astrobj{Koposov 53}.

\textbf{\astrobj{Be 89} :}   The core MF slope of \astrobj{Be 89} is relatively flat ($\chi=0.330 \pm 0.650$) with respect to outskirts of the cluster. The slope of the halo ($\chi=0.936 \pm 0.474$) and the overall ($\chi=0.789 \pm 0.522$) of \astrobj{Be 89} are not in agreement with Salpeter$^\prime$s value. 
\begin{figure}
\centering
\epsfig{file=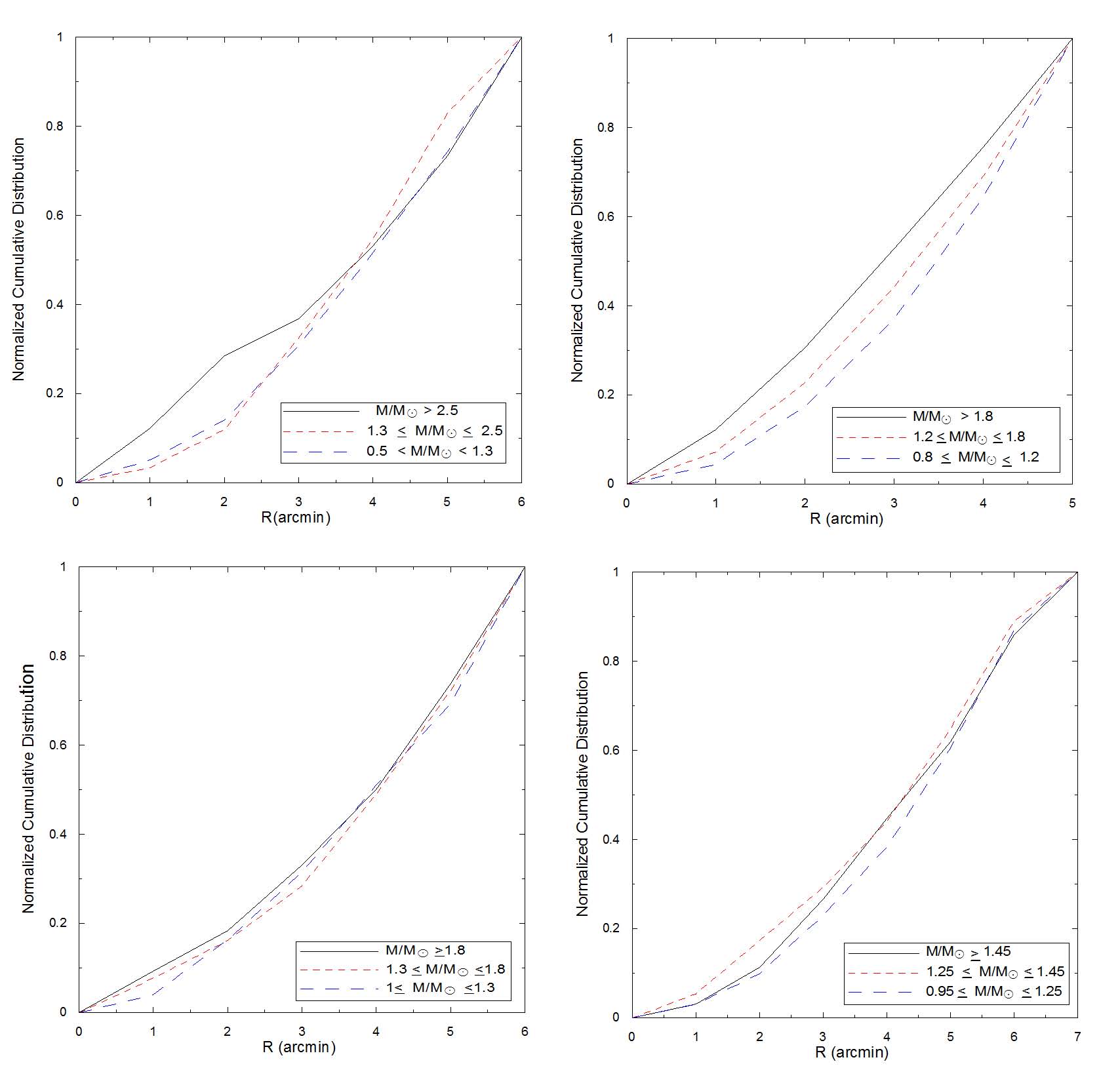, width=15cm}
\caption{Cumulated frequency distribution for the clusters. From left to right the curves represent the degree of mass segregation of stars in the cluster regions. Relatively massive stars accumulate with radius more steeply than the less massive stars do.} 
%\label{figure7}
\end{figure}
The flatten of MF slope of this cluster from the halo to the core might be comparatively explained by intrinsic dynamical evolution effects such as low mass members$^\prime$ evaporation and mass segregation with a relatively long time scale \citep{bonatto2006}. External tidal effects from bulge-disc direction, probable encounters with molecular clouds and sudden variations of interstellar gas density along orbits of cluster may have accelerated dynamic evolution of \astrobj{Be 89} due to the its Galactic location. The combined dynamical effects are expected to be significant enough to alter the mass profile in \astrobj{Be 89}, because of its old age.

\subsection{Dynamical states}
The clusters under consideration have different radial mass distributions which show various dynamical stages. It is obvious that all the samples have sustained an evident mass segregation effect which is generated by different physical processes. The derived mass ranges of the four clusters are consistent with the their nuclear ages, as expected. The youngest object, \astrobj{NGC 2571}, contains relatively much high mass stars, while the older ones, \astrobj{NGC 6802}, \astrobj{Koposov 53} and \astrobj{Be 89} contain mostly low mass stars. Our interpretations about the cluster dynamic states are based on the assumption that all stars in the limiting radii assumed as members. A more accurate determination of mass functions are needed to apply an efficient field stars decontamination procedure. The evidence for mass segregation in a cluster may also be seen from Figure 7. One can mention a decrease in the number of stars on the outer side of clusters. This is because in the formation stage, clusters may have a uniform spatial mass distribution. But after dynamical evolution, low mass stars in a cluster may possess largest random velocities trying to occupy a large volume than the high mass stars do \citep{mathieu1986}. To display mass segregation in our samples, we performed star counts on all members as a function of their masses and distances from the cluster center. This is displayed in Figure 7. In this figure, the adopted mass ranges are (M/M$_{\odot} \geq$ 2.5 -- 1.3$\leq$ M/M$_{\odot} \leq$ 2.5 -- 0.5 $\leq$ M/M$_{\odot} \leq$ 1.3), ( M/M$_{\odot} \geq$ 1.8 -- 1.2$\leq$ M/M$_{\odot} \leq$ 1.8 -- 0.8 $\leq$ M/M$_{\odot} \leq$ 1.2), ( M/M$_{\odot} \geq$ 1.8 -- 1.3$\leq$ M/M$_{\odot} \leq$ 1.8 -- 1  M/M$_{\odot} \leq$ 1.3) and (M/M$_{\odot} \geq$ 1.45 -- 1.25$\leq$ M/M$_{\odot} \leq$ 1.45 -- 0.95 $\leq$ M/M$_{\odot} \leq$ 1.25) for \astrobj{NGC 2571}, \astrobj{NGC 6802}, \astrobj{Koposov 53}  and \astrobj{Be 89}, respectively.

\subsection{The Relaxation Times of the Clusters}
The relaxation time, T$_{R}$, is the time in which the individual stars exchange their energies and velocity distributions approaching a maxwellian equilibrium. This parameter is a useful tool to interpret whether the cluster reached the dynamical relaxation or not. \citep{spitzer1971} stated that;

\begin{equation}
T_{R}=\frac{8.9 \times 10^{5} N^{\frac{1}{2}}R_{h}^{\frac{3}{2}}}{<m>^{\frac{1}{2}}\log(0.4N)}
\end{equation}
where N is the number of cluster members, $R_{h}$ is the radius of half of cluster mass in pc and $<$ m $>$ is the average mass of cluster stars in solar unit. To examine the results of relaxation times, we have re-calculated $T_{R}$ values by using the method given by \citep{binney1987}:

\begin{equation}
T_{R}=\frac{N}{8\ln{N}}t_{cr}
\end{equation}
where $t_{cr} = R/\sigma_{\nu }$ is the crossing time, N is the (total) number of stars and $\sigma_{\nu }$ is the velocity dispersion, which has been accepted usually $\approx 3 km s^{-1}$. Using this method we obtained separately  $T_{R}$ values for the overall and the core regions of the clusters as shown in Table 6. Additionally, we also derived evolutionary parameters, $\tau= Age/T_{R}$, of our samples. The comparison of different overall $T_{R}$ results are given Table 7.

\section{Conclusion}
In the present work, we analysed \astrobj{NGC 2571}, \astrobj{NGC 6802}, \astrobj{Koposov 53}  and \astrobj{Be 89}. These objects are useful for studying the disc subsystem to which the clusters belong. We have found their ages in the range of 40 Myr - 2.5 Gyr and total masses in the range of 746 - 1585 M$_\odot$.  Main conclusions are summarized as follows: 

\textbf{The astrophysical and structural parameters:} The results of the comparison of astrophysical parameters with other studies are listed in Table 2. As a result one can state that \astrobj{NGC 6802} is inside, whereas \astrobj{NGC 2571}, \astrobj{Koposov 53} and \astrobj{Be 89} are outside of the Solar circle. The reduced final reddening values from the dust maps of SDF are found to be $E(B - V)_{A}$ = 0.18, 0.72, 0.28 and 0.61 for \astrobj{NGC 2571}, \astrobj{NGC 6802}, \astrobj{Koposov 53} and \astrobj{Be89}, respectively. Our limiting and core radius estimations ($R_{lim}$ and $R_{core}$) are $(2.80, 0.84)$ pc, $(3.15, 0.54)$ pc, $(6.1,0.73)$ pc and $(5.94, 2.56)$ pc for \astrobj{NGC 2571}, \astrobj{NGC 6802}, \astrobj{Koposov 53} and \astrobj{Be 89}, respectively. $R_{lim}$ $\,$ and $\,$ $R_{core}$ values of the four clusters are underestimates due to the combinations of differential reddening and field stars contamination. These are expected to increase their efficiency of non-populous clusters projected against the dense Galactic fields. The relation between limiting and core radii of the clusters mainly depends on the Galactic location, the total mass, the number density of member stars and the time scale of pre-evolution process. The variations of last two parameters have been controlled by two independent time scales which are nuclear and dynamic scales, respectively. The masses and the integrated luminosity functions change with evolutionary states, and are mainly governed by physical conditions within the central parts of the clusters. As compared with previous works, we found differences in some parameters, such as limiting radius, reddening and age. Most of those studies were based on spatial and magnitude limited data. However, the discrepancies can be largely accounted for by the field stars contamination and different analytical methods. 

\textbf{Dynamical States:} The clusters in the present work are expected to suffer strong tidal stress which have been driven by encounters with molecular clouds and shock waves from disk and or bulge directions tending to dynamically hot clusters regions. The low mass member stars$^\prime$ segregation and the dynamical expansion processes in these clusters have enhanced due to combined external disruptive tidal effects from the Galactic plane. These destructive or/and external dynamic evolution effects are more severe for less populous clusters at large radii, especially located inside of the central parts of Galaxy. We define two main dissolving process for the outer parts of our samples: (i) Cluster expansion due to dynamical evolution; (ii) limiting radius becomes indistinguishable from background level. \astrobj{NGC 2571} presents evident sings of dynamical evolution within its limiting radius, with a relatively flat core MF slope ($\chi=0.287\pm0455$) and more steeper MF slopes ($\chi=0.439\pm0.310$, $\chi=0.432\pm0.315$) at the halo and the overall regions. \astrobj{NGC 6802} has MF slopes which are quite flat ($\chi=0.229\pm 0.111$), in the core, and very steep ($\chi=1.731\pm 0318$, $\chi=1.698\pm 0.289$) in the halo and the overall regions. The dynamic evolutions of \astrobj{NGC 6802} might have been accelerated by disruptive effects from both Galactic fields and molecular clouds, since this cluster is located at the Solar circle. \astrobj{Koposov 53} has MF slopes are slightly flat($\chi= 0.808\pm 0.241$) in the core, and relatively steep ($\chi=1.364\pm 0.411$, $\chi=1.287\pm 0.392$) in the halo and the overall regions. The variation of MF slopes from the center to outskirts might be partly explained that a mid-level mass segregation and member stars evaporation processes have occurred in \astrobj{Koposov 53} which is located at the Galactic anti-center direction. The relatively flat core MF slope ($\chi=0.330\pm 0.650$), and the more steep the halo and overall MF slopes ($\chi= 0.936\pm 0.474$, $\chi= 0.789\pm 0.522$) of \astrobj{Be 89} indicate that the variation in $\chi$ is considerably evident from the core to the external regions. We also used an evolutionary parameter ($\tau= Age/T_{R}$) which was defined by \citep{bonatto2005} as a structural indicator. In particular, significant flattening in core and overall MFs due to dynamical effects such as mass segregation is expected to occur for  $\tau_{core} \geq 100$ and $\tau_{overall}\geq 7$, respectively. The evolutionary parameters are calculated by following the approach of Bonatto and Bica (2005, 2006, 2007). It implies that the dynamical states of the clusters of interest are very important except \astrobj{NGC 2571}, because of its relatively young age. The combined effect of interstellar dust and field-star contamination may also lead to significant deviation between determined mass ranges and real ones due to faint/low masses cluster stars. Although, differential reddening and field-star contamination effect on mass distributions may lead to some difficulties for determination of dynamical properties of the clusters. We consider two probably reasons to explain relatively large core sizes of \astrobj{Be 89}, and \astrobj{NGC 2571}. These are mass-loss processes which relate to rapid stellar evolution of massive clusters stars and alternatively central black holes, formed in the supernova explosions of the most massive member stars which may lead heating the core regions of the clusters, respectively. The relatively prolonged core structures of \astrobj{Be 89} and \astrobj{NGC 2571} have been operated by different evolutionary time scales depending highly both on central densities and initial mass distributions. In the case of \astrobj{NGC 2571}, the degree of core expansion might have increased within the early evolutionary state as suggested by Goodwin and Bastian (2006) which is about 10-20 Myr for a young massive star cluster if the combined effect of massive stellar winds and supernova explosions had led to significant gas escape from the cluster.

\section{Acknowledgment}
We thank the anonymous referee for her/his helpful suggestions. 
This publication makes use of data products from the Two Micron All Sky Survey, which is a joint project of the University of Massachusetts and Infrared Processing and Analysis center / California Institute of Technology, funded by the National Aeronautics and Space Administration and the National Science Foundation. This research has made use of WEBDA database, operated at the Institute for Astronomy of University of Vienna. This work was supported by the Research Fund of the Erciyes University. Project number: FBY-10-3296.

\end{document}